# Effect of structural supermodulation on superconductivity in tri-layer cuprate $Bi_2Sr_2Ca_2Cu_3O_{10+\delta}$

Changwei Zou[1,*], Zhenqi Hao[1,*], Haiwei Li[1], Xintong Li[1], Shusen Ye[1], Li Yu[2], Chengtian Lin[3], Yayu Wang[1,4 †]

[1]*State Key Laboratory of Low Dimensional Quantum Physics, Department of Physics, Tsinghua University, Beijing 100084, P. R. China*

[2]*Beijing National Laboratory for Condensed Matter Physics, Institute of Physics, Chinese Academy of Sciences, Beijing 100190, P. R. China*

[3]*Max Planck Inst Solid State Res, Heisenbergstr 1, D-70569 Stuttgart, Germany*

[4]*Collaborative Innovation Center of Quantum Matter, Beijing, China*

*\* These authors contributed equally to this work.*

† Email: yayuwang@tsinghua.edu.cn

We investigate the spatial and doping evolutions of the superconducting properties of tri-layer cuprate $Bi_2Sr_2Ca_2Cu_3O_{10+\delta}$ by using scanning tunneling microscopy and spectroscopy. Both the superconducting coherence peak and gap size exhibit periodic variations with the structural supermodulation, but the effect is much more pronounced in the underdoped regime than at optimal doping. Moreover, a new type of tunneling spectrum characterized by two superconducting gaps emerges with increasing doping, and the two-gap features also correlate with the supermodulation. We propose that the interaction between the inequivalent outer and inner $CuO_2$ planes is responsible for these novel features that are unique to tri-layer cuprates.

Although the mechanism of high temperature superconductivity in cuprates is still under debate, there are important empirical rules that put strong constraints on the choice of theoretical models. A well-established trend is that for the same family of cuprates, the maximum transition temperature ($T_c$) increases with the number of $CuO_2$ planes in each unit cell up to the tri-layer limit [1]. The tri-layer cuprates not only have the highest $T_c$, but also have the unique situation with two types of inequivalent $CuO_2$ planes. As shown by the schematic crystal structure in Fig. 1(a), the inner $CuO_2$ plane (IP) is sandwiched between two outer planes (OP), and is expected to have distinct electronic structure due to the absence of neighboring apical oxygen on both sides. It is generally believed that the IP has lower carrier density and larger pairing potential, whereas the OP has higher carrier density and stronger phase rigidity [2]. The cooperation of these two complementary merits, made possible by the interaction between the IP and OP, is beneficial for optimizing the superconductivity [3]. However, there is still debate regarding how the two types of $CuO_2$ layers couple with each other, whether through the single particle proximity effect [4] or via the Josephson tunneling of Cooper pairs between the two superconducting (SC) layers [5].

The Bi-based tri-layer cuprate $Bi_2Sr_2Ca_2Cu_3O_{10+\delta}$ (Bi-2223) has attracted particular attention because it is readily cleavable and the charge neutral surface is ideal for surface sensitive electronic structure probes such as angle-resolved photoemission spectroscopy (ARPES) [6,7] and scanning tunneling microscopy (STM) [8,9]. ARPES experiments on Bi-2223 have revealed the existence of two split bands [10,11], presumably originated from the IP and OP respectively. Another interesting issue specific to the Bi-family cuprates is the existence of structural 'supermodulation', which causes dramatic distortions from the flat crystalline planes. Although the supermodulation is usually considered as a nuisance in data analysis, from a more optimistic perspective it provides a new knob for perturbing the $CuO_2$ planes. STM experiments on double-layer $Bi_2Sr_2CaCu_2O_{8+\delta}$ (Bi-2212) have revealed a periodic variation of the SC gap size with structural supermodulation [12], but a systematic study in Bi-2223 is still lacking. The response of Cooper pairs to the lattice distortion within each $CuO_2$ plane, plus the interaction between the inner and outer planes, may provide unique clues about superconductivity that are unavailable elsewhere.

In this letter, we use STM to investigate the effect of structural supermodulation on superconductivity in Bi-2223 tri-layer cuprates. We find that both the SC coherence peak height and gap size exhibit periodic variations with the supermodulation, but the effect is much more pronounced in the underdoped regime than at optimal doping. Moreover, a new type of spectrum characterized by two SC gaps emerges with increasing doping, and its features also correlate with the supermodulation. We propose that the intricate interaction between two inequivalent $CuO_2$ planes with distinct doping evolutions is responsible for these novel features that are unique to tri-layer cuprates.

The Bi-2223 single crystals studied here are grown by the travelling solvent floating zone method and are post annealed in mixed Ar and $O_2$ gas flow to control the doping level [13,14]. The single crystal is cleaved *in situ* at room temperature, and is then transferred into a home-built STM system working at 5 K. An electrochemically etched tungsten tip is treated and calibrated on Au(111) surface before the measurement, following the same procedure as in our previous report [15]. The $dI/dV$ differential conductance, which is approximately proportional to the local density of state (DOS), is obtained by the *ac* lock-in method with modulation frequency $f$ = 573 Hz.

Figure 1(c) displays the topographic image of an underdoped Bi-2223 with $T_c^{onset}$ ~ 110 K. Both the atomic lattice of the exposed BiO surface and the supermodulation of Bi atoms with ~2.6 nm periodicity can be clearly resolved. In a zoomed-in image in Fig. 1(b), we assign a spatial phase $\theta$ to facilitate the data analysis and discussions [12], where the supermodulation ridge and valley correspond to $\theta$ = 0° and 180°, respectively. To reveal the electronic structure of a particular relative location (or $\theta$ value), we collect 128 $dI/dV$ curves along a column of pixels parallel to the supermodulation (such as the magenta lines in Fig. 1(c)). The averaged spectra are displayed in Fig. 1(d), and a vertical offset is added to each curve for clarity. It can be immediately recognized that there exist strong periodic variations of the $dI/dV$ curves with the supermodulation. The spectra at the valley have much sharper SC coherence peak and smaller SC gap compared to that obtained at the ridge, and the evolution between them is continuous.

We have performed the same measurements on an optimally doped Bi-2223 with $T_c^{onset}$ ~ 113 K. It has been known that the $T_c$ of Bi-2223 is not very sensitive to the hole concentration [14,16], which could be due to the distinct doping evolution and interplay between the two inequivalent $CuO_2$ planes [17,18]. The topographic image in Fig. 1(e) is similar to the underdoped sample, but the spatially dependent $dI/dV$ curves in Fig. 1(f) show much weaker though discernable periodic variations with supermodulation.

The overall gap size of the optimally doped sample, defined as the energy between the Fermi level and coherence peak, is around 45 meV. It is significantly smaller than that of the underdoped sample with $\Delta \sim 67$ meV, which confirms the increase of hole density.

To directly visualize the spatial patterns of superconductivity, we perform spectroscopic imaging measurements on the underdoped Bi-2223. Figure 2(a) shows the topographic image of a large area with size $42 \times 42$ nm$^2$, on which $dI/dV$ curves are taken on a dense grid. In Fig. 2(b) and 2(c), we display the spatial information of SC gap size $\Delta(\mathbf{r})$ and coherence peak height $h(\mathbf{r})$ extracted from the spectral grid. Both quantities are shown as the relative difference to their mean values, i.e., $\frac{\delta\Delta}{\bar{\Delta}} = \frac{\Delta(\mathbf{r}) - \bar{\Delta}}{\bar{\Delta}}$ and $\frac{\delta h}{\bar{h}} = \frac{h(\mathbf{r}) - \bar{h}}{\bar{h}}$. Despite the granularity in both maps, which are typical for cuprates due to strong inhomogeneities [19,20], there are evident vertical stripy features demonstrating the periodic distribution of the two quantities. The periodicity is exactly the same as the structural supermodulation (see the Supplementary Material, Fig. S1), and the valley shows smaller SC gap (blue color in Fig. 2(b)) and sharper coherence peak (red color in Fig. 2(c)). Because $\Delta(\mathbf{r})$ indicates the local Cooper pairing strength and $h(\mathbf{r})$ is approximately proportional to the superfluid density [21], these results demonstrate that supermodulation has strong influence on the two key ingredients of superconductivity in underdoped Bi-2223. The same set of data for the optimally doped Bi-2223 are shown in Fig. 2(d) to 2(f). Although there are still faint vertical streaks in the $\Delta(\mathbf{r})$ and $h(\mathbf{r})$ maps, their correlation to the structural supermodulation is much less obvious than that in the underdoped sample.

In Fig. 3(a) we directly compare the $dI/dV$ spectra at the centers of the supermodulation ridge ($\theta = 0°$) and valley ($\theta = 180°$) averaged over the whole grid. The spectral difference in the underdoped sample is very pronounced, whereas that in the optimally doped sample is rather weak. In Fig. 3(b), the normalized gap amplitude and coherence peak height in the two Bi-2223 samples are plotted as a function of the spatial phase $\theta$. All the data can be fit by a cosine function, and the peak-to-peak amplitude of the underdoped Bi-2223 (black circles) are ~25% and ~38% for $\Delta(\mathbf{r})$ and $h(\mathbf{r})$, respectively. In contrast, in the optimally doped sample (red squares) the peak-to-peak variations of $\Delta(\mathbf{r})$ and $h(\mathbf{r})$ are merely ~5% and ~8%. For comparison, the same quantities of a slightly underdoped Bi-2212 are plotted as blue triangles (the raw data are shown in the Supplementary Material, Fig. S2). The amplitudes are ~9% and ~15%, which are consistent with that reported previously [12], and lie between the two Bi-2223 samples.

Another difference between the two Bi-2223 is the emergence of two SC gaps in the *dI/dV* spectra of the optimally doped sample, which are rarely seen in the underdoped sample. About 75% of the spectra are found to exhibit two gaps, and Fig. 4(a) shows eight representative *dI/dV* curves with two coherence peaks at each side of the Fermi level in a nearly particle-hole symmetric manner. To extract the properties of the two gaps, we fit each spectrum by using two Lorentzian functions after subtracting the background outside the coherence peaks (see the Supplementary Material, SI C for details). The histogram in Fig. 4(b) summarizes the distribution of the two gap sizes, both exhibiting Gaussian lineshapes with two peaks at 40 meV and 52 meV, respectively. These values are close to the two average gap sizes of 43 meV and 60 meV observed by ARPES in optimally doped Bi-2223 [10], so the small (large) gap can be ascribed to the OP (IP) respectively. In Fig. 4(c) we plot the fitted peak height $h(\mathbf{r})$ of each gap as a function of $\theta$, both showing a cosine dependence with the supermodulation. The OP gap has a higher coherence peak, presumably due to larger superfluid density, and it shows stronger spatial variation (peak-to-peak amplitude ~14%) than the IP (~4%).

There are several theoretical models about the effect of supermodulation on superconductivity in Bi-based cuprates. From the structural point of view, a number of factors may give rise to such phenomenon, especially the strong lattice imperfections on the ridge. First, the displacement of apical oxygen and distortion of $CuO_5$ pyramid are most severe on the ridge [22], which can cause an enhancement of SC gap size due to locally maximized paring strength [23-25]. Moreover, on the ridge the distance between the apical oxygen and the underneath Cu atom is smaller [22,26], which can enhance the pairing gap through the effect of nearest-neighbor 'super-repulsion' [27] or covalency between Cu $3d_{x2-y2}$ and O $2p_z$ orbitals [28]. The smaller distance also reduces the SC phase ordering due to delocalization of Cu 4s orbital and smaller intralayer hopping range [29]. From the charge doping point of view, the 'Bi-dilute' ridge strongly affects the Sr sites in the SrO plane, replacing $Sr^{2+}$ by $Bi^{3+}$ thus locally doping electrons into $CuO_2$ plane [22,30]. The reduced hole density at the ridge region also leads to larger SC gap size and weaker phase coherence.

However, none of these factors can tell why underdoped Bi-2223 responses much more strongly to the supermodulation than the optimally doped sample. Such doping dependence is absent in Bi-2212 [12], suggesting that it is a unique feature of tri-layer Bi-2223 due to the additional inner $CuO_2$ plane. Because there was no theoretical discussion about this behavior before, here we will follow a phenomenological path. The IP in tri-layer cuprates is known to have much lower doping level than the OP [31-34], and is most likely non-SC in the underdoped sample studied here because only one SC gap is observed. It has been proposed that the underdoped IP and overdoped OP couple with each other through the single-electron proximity effect [35,36], and induces an enhancement of SC gap in OP but a suppression of its superfluid density. Interestingly, in this picture the influence of IP on the superconductivity of OP is exactly the same as that of the supermodulation ridge. The cooperation of these two different mechanisms in underdoped tri-layer cuprate will lead to very pronounced periodic variations of the SC gap size and coherence peak with the supermodulation. The proximity effect is supported by the SC gap size of ~ 67 meV in underdoped Bi-2223, contributed by the overdoped OP, which is much larger than that in Bi-2212 with similar hole density [37].

Upon reaching the optimal doping, the IP is doped with sufficient amount of holes and becomes SC [34], as manifested by the two SC gaps in Fig. 4(a). In this regime the single-electron proximity effect between the IP and OP becomes negligible [18], and the supermodulation is the only perturbation to superconductivity. The effect is mainly acted on the OP by the neighboring apical oxygen, similar to the situation in Bi-2212. Therefore, the 14% relative variation of the OP coherence peak in Fig. 4(c) is very close to the 15% value of Bi-2212. Because the IP is farther away from the apical oxygens and is sandwiched between two outer planes, the influence of the supermodulation is significantly reduced. The IP thus lives in a relatively homogeneous environment [33], and its coherence peak variation shown in Fig. 4(c) is merely 4%. Roughly speaking, the overall response of optimally doped Bi-2223 to the supermodulation is an average of the two coexisting SC planes, and the coherence peak variation of 9% in Fig. 3(b) indeed lies between that of the two individual planes.

In conclusion, STM studies on Bi-2223 reveal the spatial and doping evolution of superconductivity, in which the underdoped regime shows much more pronounced periodic variations with supermodulation. These results can be consistently explained by considering the critical role played by the inner $CuO_2$ plane, which is unique to tri-layer cuprate. In the underdoped regime, the IP couples with the OP via single particle proximity effect, whereas in the optimally doped regime the IP couples with the OP via Josephson tunneling of Cooper pairs. The distinct doping evolution of the two inequivalent $CuO_2$ planes and the intricate interplay between them are key factors in determining the peculiar phase diagram of tri-layer cuprates, as well as the highest $T_c$.

**Acknowledgements:** This work was supported by the Basic Science Center Project of NSFC under grant No. 51788104, NSFC grant 11534007, MOST of China grant 2015CB921000, 2017YFA0302900. This work is supported in part by the Beijing Advanced Innovation Center for Future Chip (ICFC).

Figure Captions:

FIG. 1. (a) Schematic half-unit-cell crystal structure of Bi-2223. (b) Illustration of the spatial phase $\theta$ of supermodulation. (c) Topographic image (24 × 24 nm$^2$) and (d) spatial variation of tunneling spectra of an underdoped (UD) Bi-2223, where the red arrows indicate the direction along which the spectra are displayed. (e) and (f) The same plots for an optimally doped (OPT) Bi-2223 sample. The topographic images in (c) and (e) are taken with bias voltage $V$ = -250 mV and tunneling current $I$ = 5 pA.

FIG. 2. The topographic image (a), the spatial distribution of relative gap size variation (b) and coherence peak height variation (c) of the underdoped Bi-2223 extracted from the spectral grid. (d)-(f) The same set of maps in the optimally doped Bi-2223. The underdoped sample exhibits much more pronounced periodic variations with the supermodulation than the optimally doped sample.

FIG. 3. (a) Averaged $dI/dV$ curves on the supermodulation ridge and valley in underdoped (lower panel) and optimally doped (upper panel) Bi-2223, normalized to the background conductance $dI/dV$ (+135 mV). (b) Normalized gap size (lower panel) and coherence peak height (upper panel) of three different samples as a function of $\theta$. The solid symbols are experimental data and the curves are fittings by a cosine function.

FIG. 4. Emergence of two SC gaps in optimally doped Bi-2223. (a) Eight representative $dI/dV$ curves showing two SC gaps. (b) Histogram of the larger IP gap (blue) and smaller OP gap (red), which can be fit by Gaussian functions (the solid lines) that peak at $\Delta_{OP}$ = 40 meV and $\Delta_{IP}$ = 52 meV, respectively. (c) Peak heights of the fitted double Lorentzian curves as a function of $\theta$. The red and blue curves are cosine fittings of the statistical data.

Fig. 1

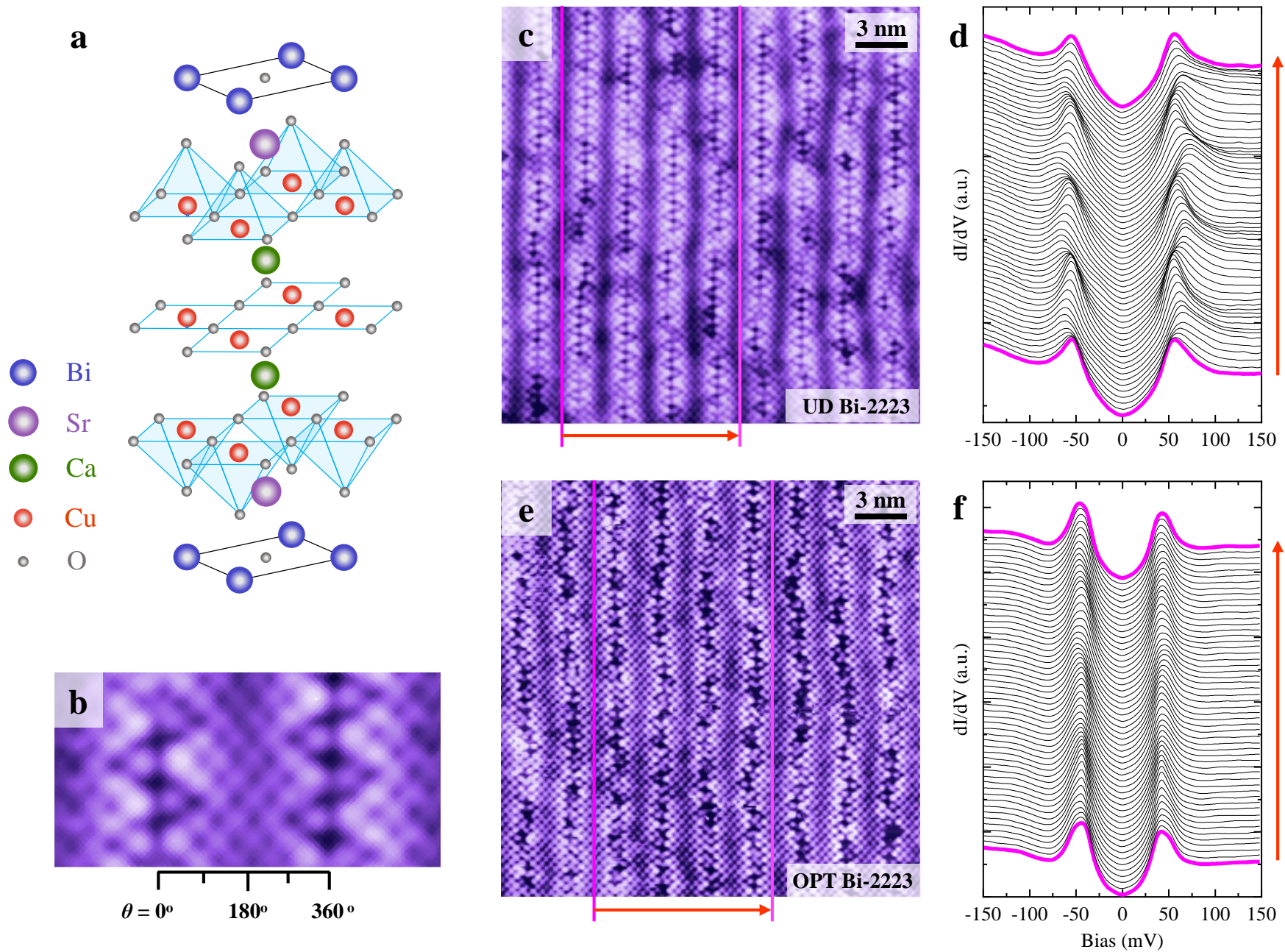

Fig. 2

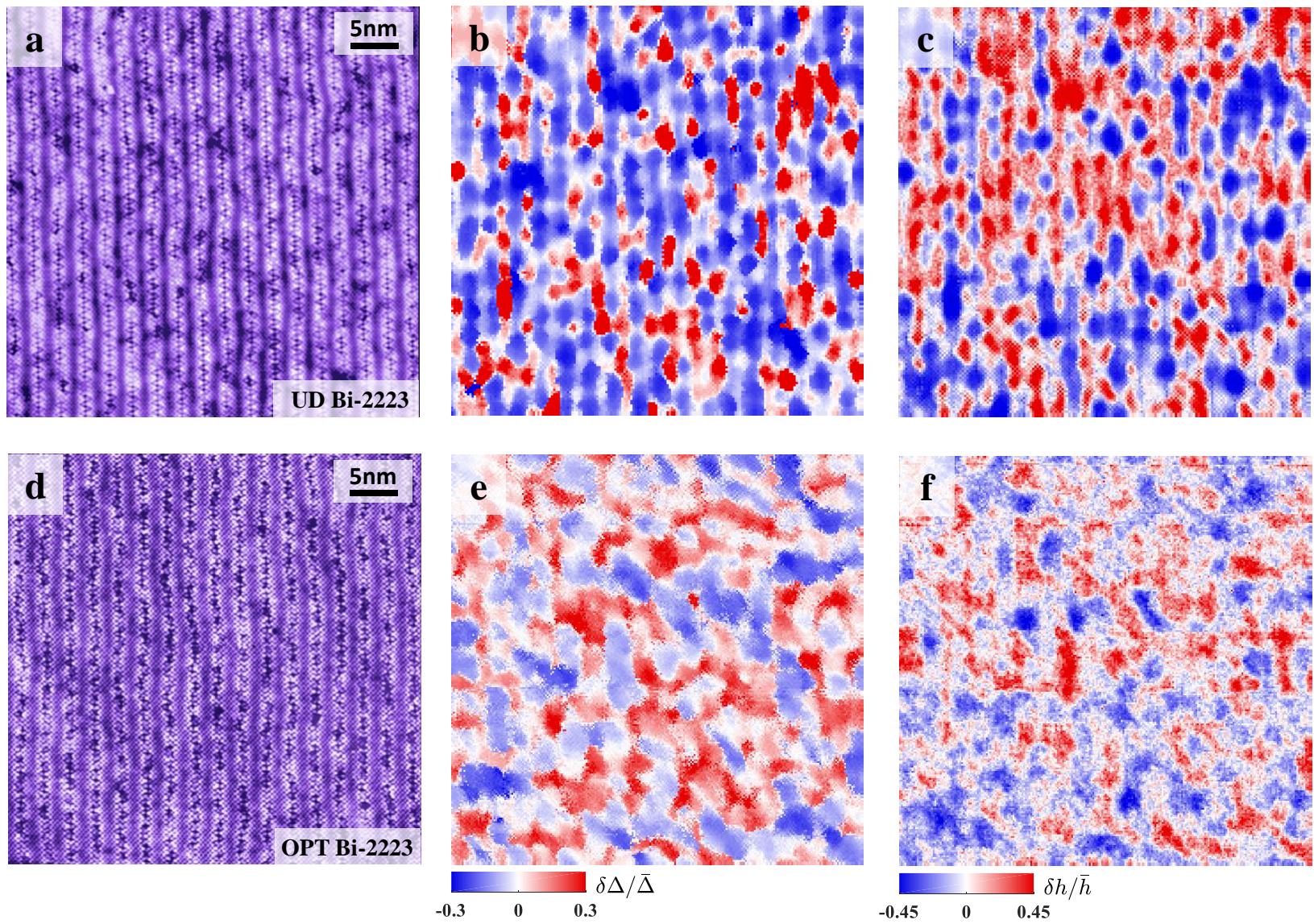

Fig. 3

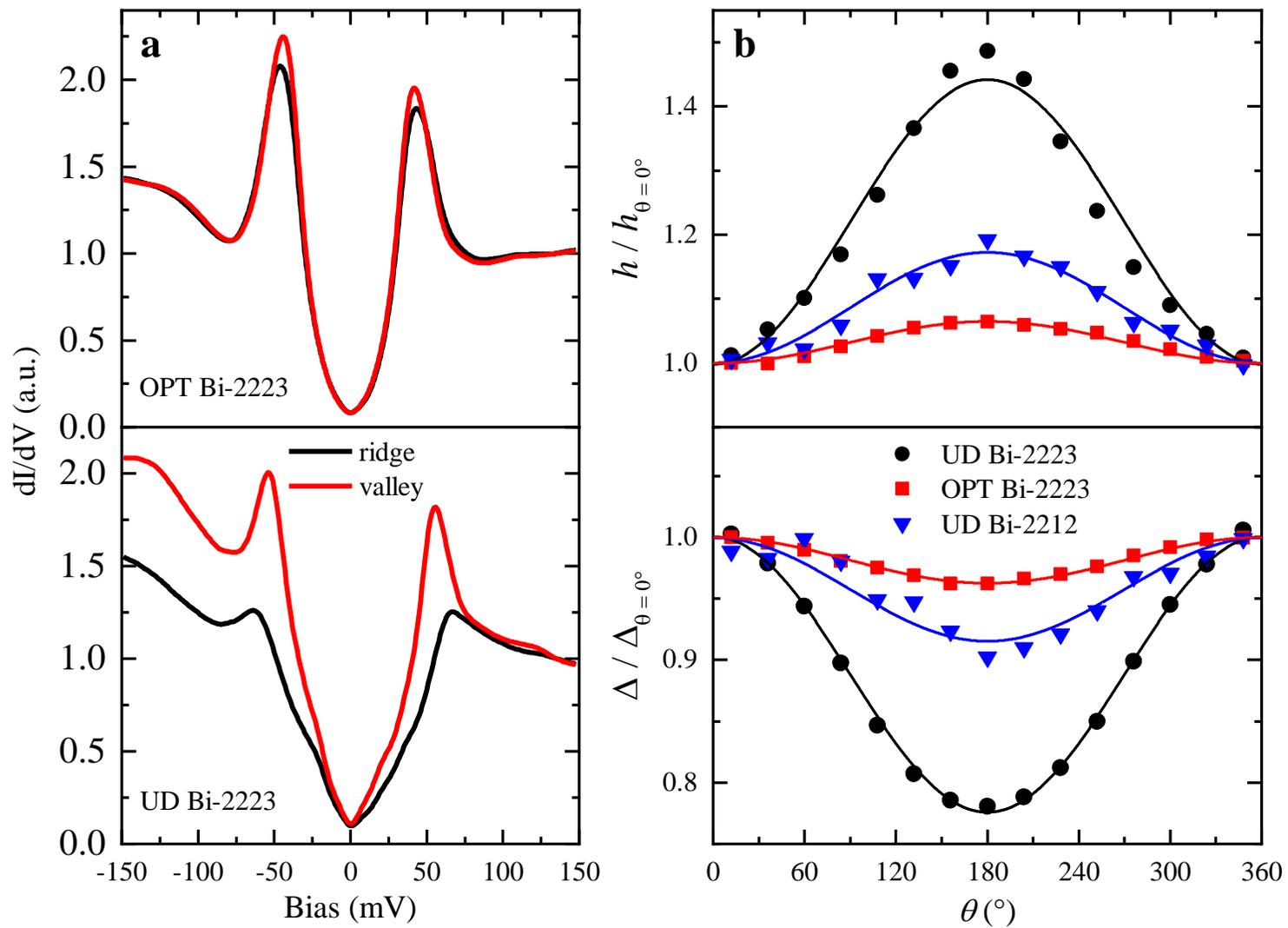

Fig. 4

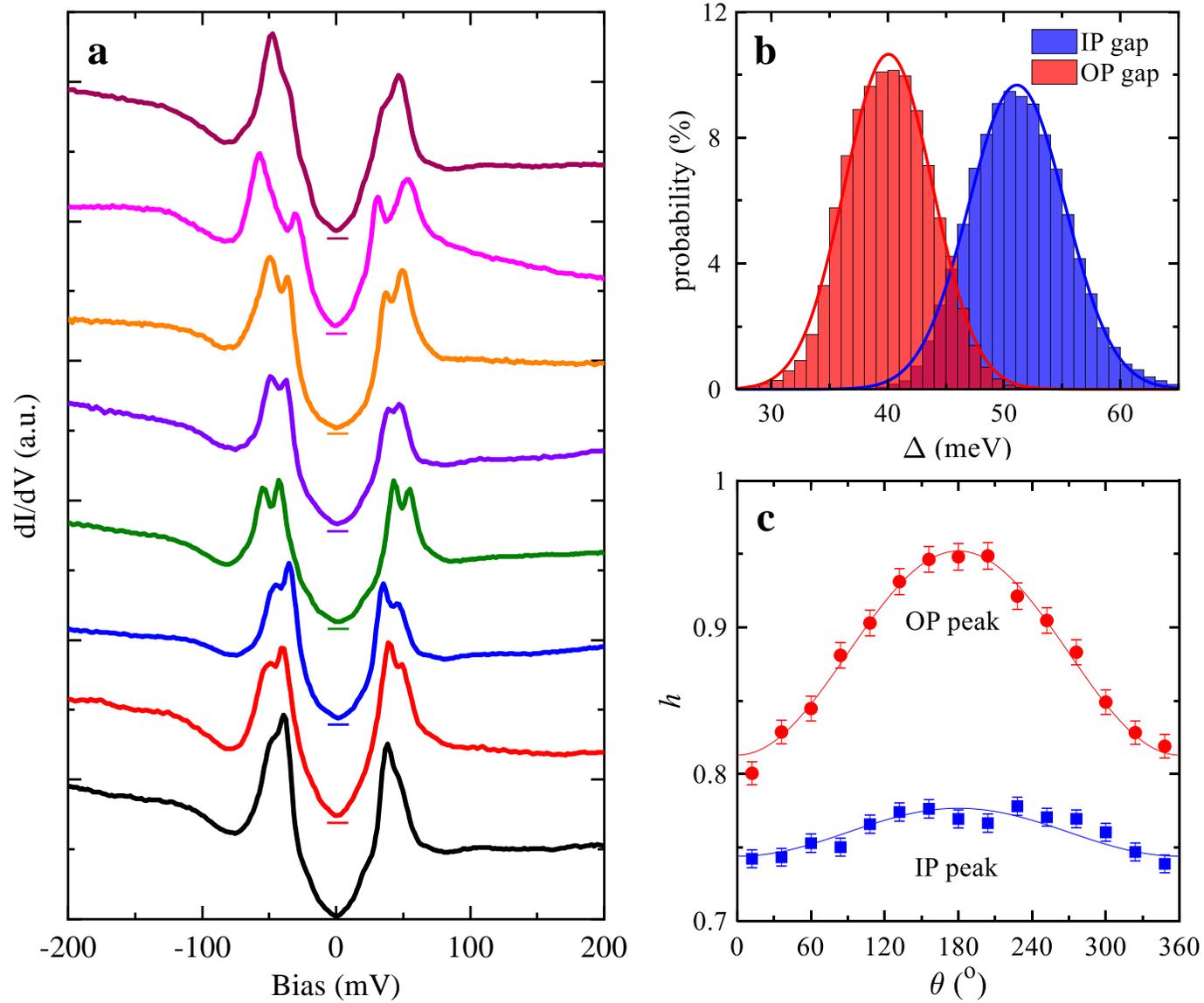